# Towards a Secure and Reliable IT-Ecosystem in Seaports1

**Tobias Brandt**
tobias.brandt@dfki.de
DFKI GmbH
Bremen, Germany

**Dieter Hutter**
hutter@dfki.de
DFKI GmbH
Bremen, Germany

**Christian Maeder**
c.maeder@uni-bremen.de
Bremen University
Bremen, Germany

**Rainer Müller**
r_mueller@isl.org
Institute of Shipping Economics and Logistics
Bremen, Germany

**Abstract**

Digitalization in seaports dovetails the IT-infrastructure of various actors (e.g., shipping companies, terminals, customs, port authorities) to process complex workflows for shipping containers. The security of these workflows relies not only on the security of each individual actor but actors must also provide additional guarantees to other actors like, for instance, respecting obligations related to received data or checking the integrity of workflows observed so far. This paper analyses global security requirements (e.g., accountability, confidentiality) of the workflows and decomposes them - according to the way workflow data is stored and distributed - into requirements and obligations for the individual actors. Security mechanisms are presented to satisfy the resulting requirements, which together with the guarantees of all individual actors will guarantee the security of the overall workflow.

*Keywords:* Port Community System, Secure workflows, Accountability, Access Control

## 1. Introduction

With the progressing digitalization, critical infrastructures, such as seaports, are increasingly becoming victims to cyberattacks. While non-targeted attacks are mainly initiated by criminal groups for extortion reasons, also state-based agencies use targeted attacks increasingly to

1 This research is supported by the German Federal Ministry of Traffic and Digital Infrastructre in the project SecProPort (Scalable Security Architectures for Commercial Workflows in German Harbors) under grant 19H18012E.

destabilize foreign countries. However, maritime transport is of central importance most industrial economies. Major disturbances of large ports could lead to tremendous negative effects on maritime supply chains and the economy. Successful cyberattacks can create significant transport delays and severe consequences for the entire economy. In ports many different actors interact in a complex IT-ecosystem that is especially vulnerable to cyberattacks since the infiltration of one actor can infect the entire eco-system as demonstrated in the successful attack[2] on a large shipping line in 2017.

The security of such an IT-ecosystem relies on the security of the involved IT-systems of the individual actors as well as on the security of the interaction of them to perform the daily port processes. In this paper we focus on this security of the interaction between the various actors in the port. The security of an individual actor's IT-system is to some extend independent of the maritime application and subject to a combination of standard IT-security measures. We consider the workflows running between the different actors (e.g., shipping lines, terminals, customs, port authority, forwarders) and present different security mechanisms to prevent potential attacks from outside or frauds from inside the IT-ecosystem. The latter threat emanates typically from former or active (frustrated) employees of actors who have access to internal knowledge of their company and running workflows. Furthermore, insiders from different companies may join forces to mount a successful attack.

The workflows running in German ports are rather old and some of them date back to the early 20th century (Hamburger Freihafen-Lagerhaus-Gesellschaft, 1938). Meanwhile, digitalization has changed the way data are exchanged between the actors however, the overall procedure of the workflows remained the same. There is a need for a common security policy governing the IT-workflow. However, the actors operate independent IT-systems. Data exchange between the actors is rather complex because of a lack of standards and different data formats, (Li, R., et al., 2010). A plethora of different protocols with their individual security characteristics (starting from email and ending with RESTful web-API) are used for communication. Data must be often converted into different formats when transported along the workflow which brings up the issue of data integrity.

In this paper we will provide ways to formulate a common security policy governing the workflows in a port. Based on such policies we develop different approaches to communicate and store data processed in such workflows. We will discuss our approach with the help of container workflows since a multitude of actors with different roles are involved in an intermodal door-2-door container transport (Wagenaar, 1992).

---

2 c.f. https://gvnshtn.com/maersk-me-notpetya/ or https://www.wired.com/story/notpetya-cyberattack-ukraine-russia-code-crashed-the-world/

## 2. Workflows in Ports

To ensure the smooth flow of cargo through seaports, the logistical flow of goods and the information flow must be optimized. For the last point, electronic data transmission systems for ports are often used. These systems are commonly known as Port Community Systems (PCS). PCSs are centralized information and data hubs for ports supporting the integration and distribution of information from various sources. PCSs can act as a communication system to synchronize data and status of the workflow between the various actors.

PCSs provide interfaces to the systems of companies, public organizations and authorities involved in maritime transportation, such as shipping lines, freight forwarders, terminal operators, carriers (ocean, road, rail, and inland waterway), and authorities such as customs and port authorities.

A PCS enforces some legal regulations and offers other services that may cost extra fees. However, PCSs are basically only an essential communication hub for shipping lines and terminals. Although the PCS also communicates with customs, the individual clients (exporters and importers) still need to issue their declarations and transport orders directly with customs and shipping lines. Consequently, a PCS is just one - though big - player within the port ecosystem communicating via peer-to-peer (P2P). PCSs inform affected actors actively about events or state changes.

Figure 2-1 describes the process of a container export, featuring several actors and their communication via the PCS. After the container has been delivered to the terminal, the PCS is informed via an ICU (Content: container arrived at terminal) message and the shipping line via a CODECO (Content: container arrived at terminal) message. The port authority is informed if the container contains dangerous goods. The container is then stored on the terminal premises, and the PCS is informed of the container location by an LCU (Content: container was moved) message. If dangerous goods are being transported, the port authority is also informed of the new container location. Several messages are exchanged between PCS, shipping line and customs, so that custom authorities can finally release the container for export (clearance state). This clearance state is forwarded by the PCS to the terminal and the shipping line.

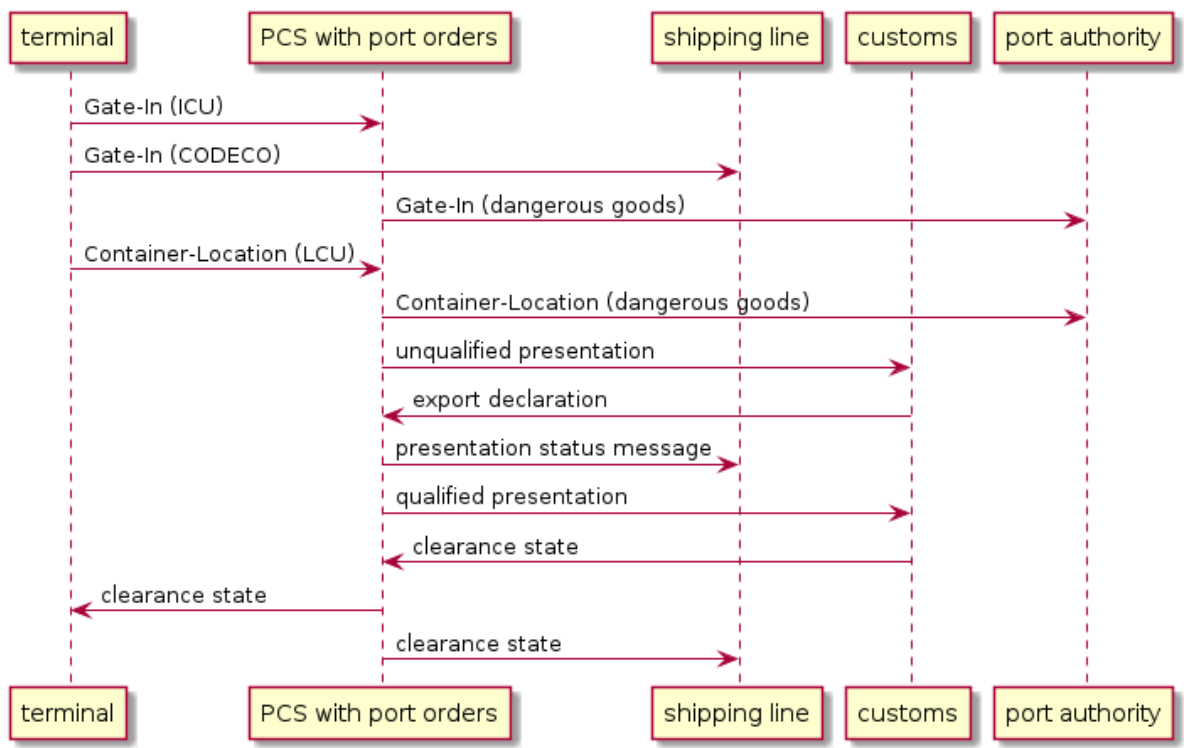

**Figure 2-1: Container Export Scenario**

A typical import process is given in Figure 2-2. During the sailing of the vessel, the shipping line transmits the import manifest via IFTMCS (contains Bill of Lading B/L) message to the PCS, with the data of the containers to be unloaded at the terminal. After that, a new port order is generated by the PCS and transmitted to the terminal. In addition, the port order is also sent to the port authority in case of dangerous goods. Afterwards, the PCS sends the import manifest to customs so that they can carry out an import risk assessment. If the container is allowed to be imported, customs send the ATB number (import customs reference) to the PCS. Afterwards, the PCS forwards the ATB number to the terminal and the shipping line by using an IFSTA (Customs status message) message.

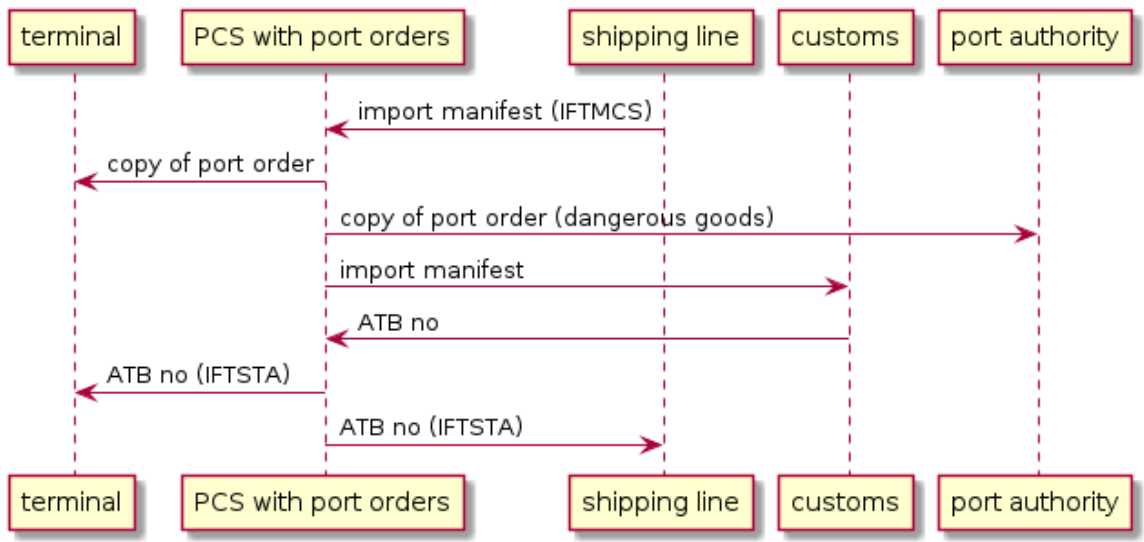

Figure 2-2: Initial Import Messages

## 3. Security Policies for Seaports

In general, a security policy regulates all activities of a system prescribing which operations on which objects are allowed for all the different subjects. Here we are only concerned with regulations for the (digital) exchange of data between different actors. Examples of such messages are IFTMCS or IFSTA messages as shown in Figure 2-2. A common format for such messages is the UN/EDIFACT standard for electronic data interchange for administration, commerce and transport (ISO 9735). Roughly speaking, this standard defines rules to syntactically structure the data used in the area resulting in an EDIFACT message as a list (or tree, respectively) of attribute-value pairs. In the following, we will adopt this approach and assume that the messages between actors of a workflow are always structured as attribute-value pairs.

The IFTMCS message provides, for instance, data containing the following attributes:

- Container number ($CNT\ No$)
- Booking number ($B\ No$)
- Number of the bill of lading ($B/L\ No$)
- Container Content description ($CNT\ C$)
- Container weight ($CNT\ W$)
- Name and address of consignor and consignee ($CSG\ DATA$)

These individual attributes are subject to different security requirements. For instance, the name and address of consignor and consignee of a container are personal data which in Europe are subject to the regulations of the General Data Protection Regulation (GDPR). Hence, they must be kept secret from unauthorized actors. The container content, for instance, may only be changed by the importer of the container (at the beginning of the workflow). Also, the various numbers used to identify the imported container or the corresponding Bill of Lading must not be changed by an unauthorized actor to avoid any confusion. The weight of a container is sensitive with respect to changes as an incorrect weight can cause an unsafe arrangement of the containers on the ship. Similar requirements hold for customs clearance.

On the one hand, messages contain data about the entities of the workflow that are subject to confidentiality or integrity conditions. On the other hand, messages represent the results of actors' decisions that determine the next steps in the workflow to be done. The integrity of such decisions (both, with respect to the decision itself and its author) is a fundamental necessity to guarantee a correct operation of the port ecosystem.

In essence, a port-wide security policy must fix the rights of the individual roles on the various types of data covered by the workflow since actors must provide also guarantees to other actors like, for instance, respecting obligations related to received data or checking the integrity of the total data received in a workflow run. In the first place, we use (role-based and attribute-based) access control to determine which actor can create or change respective data and which actor may read such data.

For example, given the attributes mentioned above, we can formalize the following access control policy (shown in Table 1) to control the distribution and manipulation of these attributes according to the needs described above. Notice that in this example we assume that the shipping line is also the freight forwarder.

Table 1: Global Access Control Policy

|  | B No. | B//L No. | CNT C | CNT W | CSG DATA | CNT No |
|---|---|---|---|---|---|---|
| **Importer** | R | R | RW | RW | RW | R |
| **Shipping Line** | RW | RW | R | RW | R | RW |
| **PCS** | R | R | - | R | - | R |
| **Terminal** | R | R | - | R | - | R |
| **Customs** | - | R | R | R | R | R |

The importer contacts a shipping line to transport the freight in a container to a port and thus he alone determines the container content and the details of consignor and consignee. The weight of the container may be also recalculated by the shipping line. While the weight is visible to all participating actors, the content as well as the personal data are only readable for the importer himself, the shipping line and customs.

Summing up, read-access permissions determine the confidentiality of attribute values occurring in a workflow run. The write-access permissions specify the integrity conditions of

IAME 2021 Conference, Rotterdam, Netherlands                                                                                     6

these values since they define the set of actors who are only allowed to create or edit individual attributes.

**4. Security Mechanisms**

In this section we present different security mechanisms to enforce the regulations given by access-control policies like the one presented in Table 1. We present two alternative solutions to guarantee the security of a composed workflow:

In the first approach, we modify the current workflow by introducing a common storage in which all workflow relevant data are stored and retrieved. Then, the security policy is enforced when accessing this storage. Our approach uses blockchain techniques to implement the data storage to allow for ecosystems in which there is no actor available who can act as a global trustworthy intermediary.

In the second approach, we keep the workflow as it is and supplement the communication between the actors with additional security measures to allow the individual actors to assess the correctness and the integrity of a workflow run (peer-to-peer communication). In essence, the access control policy is enforced by the individual actors which are provided with sufficient information to localize the origin of a fraud.

A prerequisite for the following security mechanisms is a public key infrastructure (PKI) based on root certificates that all actors trust. A PKI provides key pairs for all participating entities. A key pair consists of a private key and a public key. The public key is shared to the public and uniquely identifies an entity (person, device, etc.) possessing exclusively the corresponding private key . The private key can be used for signing data and others can verify such signatures using the corresponding public key. Special entities of a PKI are certificate authorities (CA) for digitally signing subordinate certificates. A certificate associates at least a name of an entity to a public key. The signing CA must ensure that an entity is named correctly, and that this entity controls the matching private key. CAs keep lists of issued and revoked certificates and form a hierarchy with a trusted root CA at top. Usually, even small organizations may be subordinate certificate authorities that issue digital certificates for their employees or devices adhering to regulations of a superordinate certificate authority. An obvious choice for a PKI is the secure identity management proposed by the Maritime Connectivity Platform[3] (MCP) supporting the various access control concepts like roles and attributes as well as delegation.

*4.1   Communication via a blockchain*

One way of exchanging information is to use a common datastore. All data is stored in a central place (as long as needed) and everybody (with sufficient privileges) can retrieve needed data.

---

[3] https://maritimeconnectivity.net/

In the current setting, a PCS already partly plays the role of a trusted agent providing such a central datastore.

As a central datastore for communications, a possibly distributed database may be considered. Such a database needs to be maintained and access must be controlled for those with legitimate interest. As a trusted player a PCS could be the provider of such a database. Other big and trusted players like shipping lines or terminals might maintain parts of the distributed or replicated database.

In the sequel we describe how to use a permissioned blockchain for communication. *Permissioned* means that the blockchain is not public and all participants are registered and part of a common PKI. Via digital signatures all blockchain transactions will be traceable, tamper-resistant, and accountable to known, identified participants. In contrast to a public blockchain (like Bitcoin), consensus can be achieved much more efficient without proof-of-work (PoW). By signed hash codes of preceding blocks, blockchains are immutable. Blockchains persist as many identical copies including the whole history starting from an initial block.

### 4.1.1 Permissioned Hyperledger Fabric Blockchain

As a choice for a blockchain solution we consider Hyperledger Fabric (HF) (Androulaki et al., 2018). HF is a modular and extensible open-source development platform for permissioned blockchains with pluggable consensus models. HF is supported by the Linux Foundation and IBM and is the basis of existing blockchain solutions like i.e., Tradelens[4].

When switching from P2P communication to using a blockchain, the workflow will change as depicted in Figure 4-1 for the container export scenario. Messages about containers are not directly sent to appropriate recipients but changes of container states are stored on the blockchain, and the legitimate actors are expected to read current container states and act accordingly.

For simplicity we restrict this presentation to a single container blockchain. As not messages but assets with their state changes are stored, many more blockchains for i.e., goods, packaging, customs declarations, port orders, ships, travels, etc. are conceivable.

The advantage of the blockchain is that a state change is unique and transparent as soon as the block has been committed by consensus, although this change may be visible only after some short latency by the blockchain net.

---

4 https://www.tradelens.com/

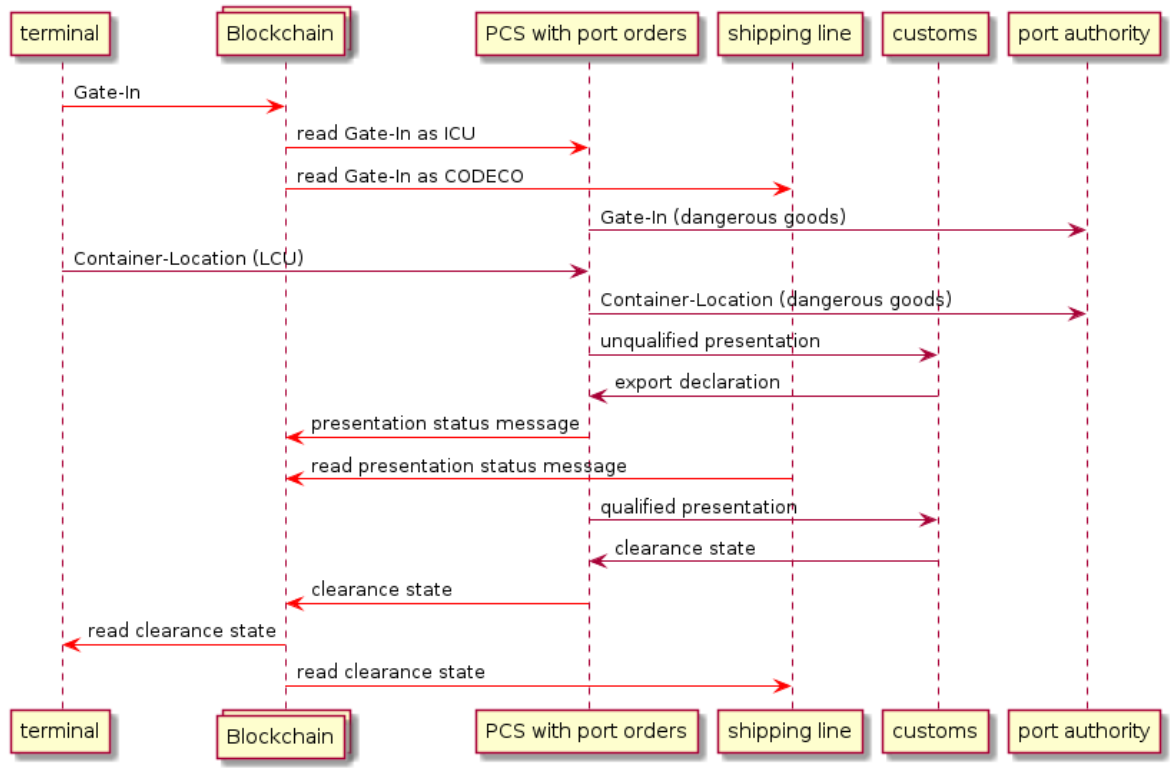

**Figure 4-1: Container Export Workflow with a Blockchain**

### *4.1.2 Enforcement of Access Control*

A container blockchain within a port ecosystem would store all registered containers from different clients, shipping lines, terminals, etc. and all these participants would run up-to-date copies of these blockchains. Analogously to (distributed) databases, access must be controlled explicitly. For the blockchain, smart contracts (also known as chaincode) need to be implemented to control access and ensure confidentiality. With HF all transactions on blockchains will be created by chaincode. External applications can only trigger transactions via chaincode. Therefore, chaincode will be the policy decision and enforcement point.

For fine grained access control we will consider role-based (Sandhu, R. S. et al., 1996) and attribute-based access control (Hu, V. C. et al., 2015), which means that some privileges are bound to the role of an organization and other privileges depend on data attributes. For instance, shipping lines may add new containers to the blockchain, whereas terminals and a PCS may only change the state of containers. During a shipment, a container goes through a lifecycle as depicted in Figure 4-2. The terminal would change the state of a new container from *created* to *delivered* as part of the *Gate-In* action or from *cleared* to *loaded* during stowage. The PCS would present a container to customs and customs would eventually clear such a container (possibly after inspections). Roles are bound to organizations, and these are part of authenticating certificates. It makes sense to provide different client applications for different roles, but access will be enforced by the chaincode.

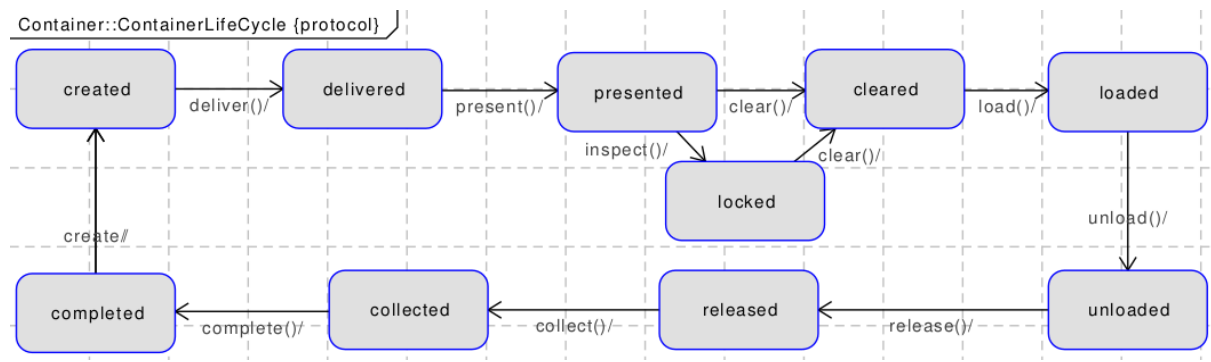

Figure 4-2: Container Lifecycle States

Role-based access control alone is not fine grained enough. Organizations are only supposed to change or see those containers for which they are in charge of. This is called *multitenancy*[5] *(Maeder, C. et al. 2020)* and requires attribute-based access control. Together with a container, also the organizations in charge of this container must be stored as an attribute (and access to this attribute must also be controlled).

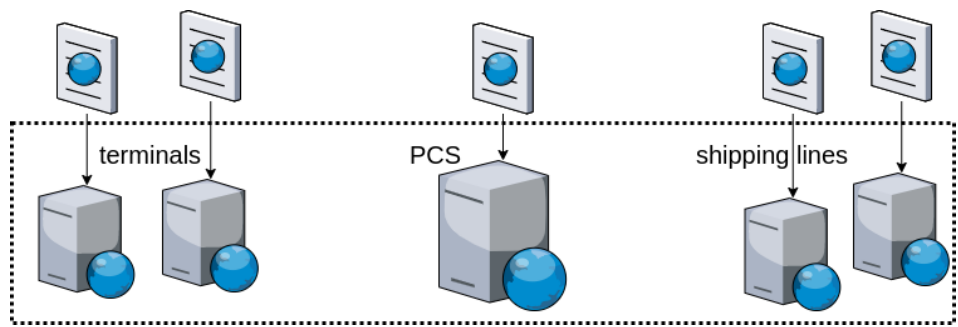

Figure 4-3: Exemplary Blockchain Net

To illustrate our blockchain access control, Figure 4-3 depicts a prototypical blockchain network of a consortium with five organizations but only three different roles. One organization plays the unique role of a PCS. The two other roles are those of shipping lines and terminals. We assume that several shipping lines and several terminals (for example, two each in Figure 4-3) communicate via a single PCS as this is very close to the current situation in German seaports. All five organization run peers hosting (a copy of) the whole container blockchain. The role specific activities are the following:

---

[5] The usage of the term multitenancy here relates to the access separation within a shared application and is not just an optimization of a web service to serve multiple tenants by one application instead of providing disjoint applications for each tenant in the first place.

- A shipping line creates a new container and chooses to which terminal this container will be delivered. The chosen terminal will likely endorse the transaction initiated by the shipping line.
- Once the container reaches the chosen terminal, the terminal acknowledges this delivery and updates the corresponding container state. This transaction can either be endorsed by the shipping line or the PCS.
- The PCS presents a delivered container to customs in order to obtain a clearance. Since customs is not part of our exemplary blockchain, the PCS will eventually set the container states to *cleared* on behalf of customs. The terminal will endorse these state changes as containers without clearance must not be loaded. In fact, PCS and customs are viewed here as a single player, as it is irrelevant, who and how containers are cleared as long as it happens according to the legal regulations.

This part of the short story describes role specific transactions. The mentioned endorsements are only one way to achieve consensus for new blocks that contain the transactions. The required number of endorsements and even the whole consensus model could be configured. Another consensus model could be Proof-by-Authority, where i.e., the PCS (or customs) would control all transactions.

Endorsements and consensus are automatic and digitally signed tasks done within milliseconds for immediate finality. Humans can only, after being authenticated, initiate digitally signed transactions using applications that trigger the corresponding chaincode. In HF, the responsibility for collecting endorsements from other peers is left to the initiating peer in order to simplify the subsequent consensus that basically only checks and orders incoming transactions. Therefore, the nodes for consensus in HF are called orderers and these nodes are run by major or even independent organizations that establish the whole blockchain net.

Typically, every organization in a blockchain consortium runs its own intermediate certificate authority (CA) to be able to create (and revoke) certificates for their employees in charge. In this simple case the mere certificate chain determines the role and access rights for the tenant of a certificate. In other cases, roles or attributes can be additional parts of certificates.

The blockchain itself can be viewed as an attribute-value store, in fact, peers keep redundant up-to-date copies of such a store to avoid the reconstruction of states by replaying the state changes stored on the blockchain. For containers we can use the world-wide unique container numbers as keys. As value of a container its mere lifecycle state is not enough to store, but additional attributes are required. Apart from their roles, different shipping lines and different terminals have additional, namely attribute-based, access restrictions. As container attributes we store at least the following two additional ones: the shipping line that *created* the container and the terminal that is supposed to handle, i.e., load or unload, the container. When a shipping line creates a container, the attribute value for the shipping line is automatically set to the creating shipping line, but this shipping line can choose any value for the terminal attribute. No other organization, including other shipping lines, can change these attribute values. Terminals can only change (and see) the states of containers with a matching terminal attribute.

The read rights are also supposed to be restricted. Shipping lines do not need to see containers of other shipping lines and even the PCS only needs to see containers within terminals. The current state of container visibility is quite permissive. Existing container tracking services allow one to find out which ship is currently carrying specific containers. Container numbers are written on the containers themselves. Containers with dangerous goods need to clearly indicate the hazard classes for emergency cases. Yet, it is a difference to observe a couple of containers on rail or roads near terminals or to read out this information from a blockchain or another central store for many containers, automatically.

### *4.1.3 Results and Discussion*

As part of ongoing work, a sample container blockchain using HF has been developed to demonstrate the support for role-based access control and multitenancy. Although the scalability and reliability of such a solution cannot be shown, the functioning can be expected to work on large scale, i.e., for hundreds of shipping lines or terminals, due to existing other blockchains like Bitcoin or Tradelens. However, this blockchain approach has also a couple of issues:

- Currently, there is no adequate support for a descriptive specification of access control. This is complicated by the fact that the current standard formalism for attribute-based access control, namely XACML (Rissanen, E., 2017), seems to be inappropriate for describing i.e., multitenancy. Hard coding access control into smart contracts is no good way to maintain an access policy reliably, as it would require code inspections by especially skilled programmers.

- From an economic viewpoint, the huge number of replicated copies of the blockchain is unattractive. For mere backup purpose a few reliable copies would be better suited. Compared with a distributed database, only a subset of the required data would be replicated temporarily as needed.

- While traceability is an important property, the inability to ever delete old data may become a problem in the long run. A solution for this would be to periodically establish new blockchains with a genesis block describing the state of a more recent past.

- Considering a consortium of a permissioned blockchain around a single PCS for a local port area, there is still a need to collaborate with other ports. Ideally, all ports (and

- PCSs) would share the same blockchain, otherwise shipping lines would need to become members of many consortia for all ports they want to serve. Different blockchains may store information, i.e., about containers, differently. If containers would need to be added to several blockchains without automatic conversions, global inconsistencies may arise.

- Since most organizations run peers that hold the entire blockchain, it may be difficult to prevent administrators of such peers from extracting information directly from the chain, thereby breaking privacy laws and undermining any multitenancy policy.

Summarizing, the transition from the current situation to a more secure/integer workflow using blockchain communication is not an easy task. However, a migration strategy would be to replace already existing traditional databases by blockchains, i.e., for containers or ships with less critical privacy requirements, and to adjust the involved workflows accordingly.

### *4.2  Peer-to-peer communication (P2P)*

In today's practice, workflows are executed in a decentralized way. There is no common datastore that represent the actual global state of a workflow instance, but all necessary information is distributed to the relevant actors participating in the workflow instance. The progress of the (virtual) workflow steps is realized by individual messages exchanged between the various actors.

In Section 3 we introduced an access-control policy governing the allowed information flow in the port ecosystem and the problem occurs of how we can enforce such a policy when all data is communicated between the actors on a peer-to-peer basis. While usually (mandatory) access control restrictions are implemented with the help of system-wide policy decision and enforcement points, there are no such central transfer points allowing us to control the traffic in the current port ecosystem. The only way to enforce such an access-control policy is that each actor must take his share to enforce it. He must control incoming messages and stop their processing if they do not satisfy his expectations as well as take specific measures when formulating outgoing messages. Analogously to security protocols, the used workflow protocol must allow an actor to detect any fake operation from which he is affected. An actor should always be recognized and held accountable if he neglects or violates the regulations.

Concerning the enforcement of write permissions, communicated data are only acceptable if they are annotated by certificates signed by actors who are authorized to write these data. Furthermore, the data must be demonstrably generated for the specific purpose under consideration and different incoming messages must agree in common attributes.

Concerning the enforcement of read permissions there is no way to prevent an evil actor to disclose sensitive data to an outsider (verbally or outside the workflow). Obviously, a trusted actor must not send any sensitive data either to unauthorized actors or to authorized actors in an unprotected way.

In the following paragraphs we illustrate our techniques to enforce these access restrictions in terms of integrity and confidentiality conditions on the communication between the actors in more detail.

### *4.2.1 Integrity*

The classical means to ensure integrity is the use of (digital) signatures. Given some data $m$ and a cryptographic hash function $h$ the hash value $h(m)$ is computed and encrypted[6] with the private key $k_a^{-1}$ of the data supplier $a$. Notice that since the corresponding decryption key $k_a$ is publicly available, everybody can decrypt this message. Both, original message and its encrypted hash value is sent to the designated receiver. Knowing $k_a^{-1}(h(m))$ and $m$, the receiver can compute $h(m)$ in two different ways: either he decrypts the encrypted hash value using the public key $k_a$ of the actor $a$, or he applies the hash value to $m$. Both computations must provide the same value otherwise either the message or its signature has been tampered with. Due to the Preimage Resistance property[7] of the cryptographic hash function $h$, an attacker with reasonable computational power is unable to compute an alternative message $m'$ that provides the same hash value as $m$. Hence, in practice any change of $m$ will invalidate its signature.

Now consider, for example, the IFTMCS message mentioned in Section 3. In particular, it contains information, supplied by the importer, about the content and the personal data of the sender and designated receiver of the container as well as some unique booking number. The shipping line will only forward this information to customs via the PCS. It is neither responsible for its correctness (e.g., the content of the container) nor is it allowed to modify the data. A signature of the data signed by the shipping line is of little help since it only guarantees the integrity of the data between shipping line and PCS. Thus, a malicious shipping line may still fake the original data of the importer before sending it to the PCS. Instead, there is a need to provide a corresponding certificate of the importer to the PCS (and later to customs). This allows, in turn, the customs to hold the importer accountable if the real content of the container does not match the declared content. Besides the importer's data, the IFTMCS message contains also data compiled by the shipping line, like for instance, the Bill of Lading or the container number. For these data, the shipping line has to provide evidence of authorship in form of a signature. Since these data are inextricably connected to the importer's data, we must include them in some way when constructing the signature of the shipping line.

---

6 By abuse of notation we will write $k(m)$ to denote the result of encrypting (or decrypting) a term $m$ by a key $k$.
7 The Preimage Resistance property states that given some value t it is practically infeasable to find a text m with $t = h(m)$.

Hence, there is a need for a fine granular signature scheme allowing one to ensure the integrity of a message by a set of signatures issued by the associated authors of the embedded data. To prevent an attacker from extracting signed partial messages from previous communications and reassemble them to now correctly signed fake messages, there is also a need to "tie" these signatures in some way together.

Since messages are lists of attribute-value pairs, in our approach a signature will protect the values $v_1 \dots v_k$ of a list of (some) attributes $n_1 \dots n_k$ occurring in a message. To compute such a signature, we concatenate the hashed values of each $v_i$, hash the result again and encrypt it with the private key $k_a^{-1}$ of the signer $a$ resulting in the signature $k_a^{-1}\left(h\big(h(v_1) \cdot \dots \cdot h(v_k)\big)\right)$ for an attribute list $n_1 \dots n_k$. The receiver of a message can check the integrity of the values of the covered attributes by comparing the value obtained by decrypting the signature (using the public key of the importer) with the result of computing the double hash of the values by herself. Notice that to compute the double hash value[8], an agent needs to know for each attribute $n_i$ either its value $v_i$ or its hashed value $h(v_i)$.

In our example, the importer signs booking number ($B\ No$), content ($CNT\ C$) and personal data ($CSG\ Data$) by the signature $k_{Importer}^{-1}(h(h(B\ No) \cdot h(CNT\ C) \cdot h(CSG\ DATA)))$. For the IFTMCS message, the shipping line will sign itself booking number ($B\ No$), bill of lading number (B/L No), container number ($CNT\ No$) and container weight ($CNT\ W$). Together with the IFTMCS message, the shipping line provides both, the signature of the importer and its own, to the PCS. Both signatures together cover all attributes of the IFTMCS message as specified in Section 3. Additionally, each of the signatures cover the booking number which links them together and prevents the shipping line from issuing another IFTMCS message with another container number but same importer data. Nevertheless, the shipping line must trust the importer (or check by itself) that he does not use the same booking number (representing a *nonce* in terms of security protocols) for different workflow instances. While multiple signatures (of different actors) provided with a message may cover a particular attribute (like the booking number in our example), at least one of them must be issued by an actor possessing write permission for this attribute to ensure integrity (which in the example is the case for the importer).

### 4.2.2 Confidentiality

When sending the IFTMCS message to the PCS, the shipping line runs into the problem that it must provide the importer's data ($CNT\ C$ and $CSG\ DATA$) to allow the PCS to forward them to customs and to verify the importer's signature. However, according to our access control matrix, the PCS has no read access to these data and providing the PCS with this data would be a violation of the common security policy. To verify the importer's signature, a simple solution would be to send the PCS neither $CNT\ C$ nor $CSG\ DATA$ but only their hashed values $h(CNT\ C)$ and $h(CSG\ DATA)$. As mentioned above, the hashed values are also sufficient to verify the signature due to the double hashing approach and the hashed values do not provide any (practically computable) information on the values themselves. However, forwarding the hashed values $h(CNT\ C)$ and $h(CSG\ DATA)$ to customs is not very helpful as they do need the real data. To keep the data secret from PCS, the shipping company must encrypt the data such that only customs is able to decrypt them. Possible solutions are either using the customs' public key directly which is inefficient or to use a (freshly generated) symmetric key $k$ and to encrypt only this key with customs' public key as it is common practice. The encrypted symmetric key will be attached to the message or can be sent separately. In our approach the shipping line will, for instance, send a pair $\langle\ h(CNT\ C),\ k(CNT\ C)\ \rangle$ consisting of hashed data and data encrypted with $k$ as the value for the attribute $CNT\ C$ to the PCS. This allows the PCS to verify the importer's signature but also customs to decrypt the encrypted values and to verify the resulting values with the help of the included hashed values.

---

[8] The double hashing technique was first used in (Mastercard and Visa, 1997) to separate payment and order information in the security protocol SET for online-shopping. Later, (Bella et al., 2005) formally verified various security properties of this protocol with the help of an interactive theorem prover.

In the final step, the PCS sends customs the import manifest message, which consists (in our simple example) of the same attributes as the IFTMCS but replaces the value of the booking number by its hashed value $h(B\ No)$ since customs has no permission to read $B\ No$ but it needs the (hashed) value to verify the importer's signature.

Notice that the confidentiality of attributes can of course only be guaranteed assuming that each actor with read access to a value does not forward it voluntarily to some unauthorized actor via some unsupervised communication line.

Summing up, for P2P communication we assume that messages are lists of attribute-value pairs. We introduce signatures to ensure the integrity of subsets of these pairs and demand that each value is covered by a signature issued by an actor who has the write permission for the respective attribute. The signature scheme allows one to use a plaintext value, its hash value or a pair of hash value and encrypted value interchangeably when validating a signature. This allows actors to verify the integrity of a message although they may not possess the necessary rights to access the (plaintext) data encoded in the message.

### 4.2.3 Realization

In order to support the migration towards a more secure workflow without compromising its availability, our proposed architecture is a mere extension of the already existing communication at the ports. Messages send or received according to the established business logic, typically EDIFACT messages, will be replaced by messages secured by signatures and encryption of message parts as described above.

Figure 4-4 shows an abstract overview of the modules and their relation to the internal components and to the outside. The complete security extension is a composition of three different software sub-modules and two digital storage systems. In the following we will describe this *Message Adapter* and its components.

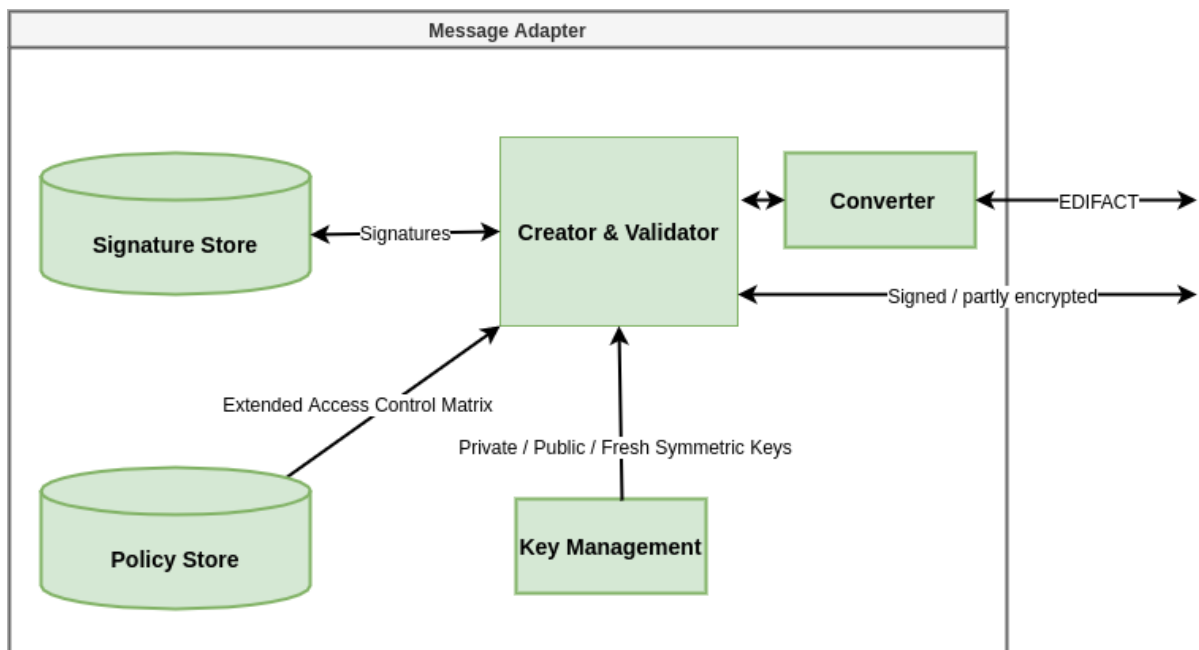

**Figure 4-4: Abstract view of the proposed modules and their interfaces**

The main component is the *Creator & Validator* module. This module has a direct communication interface to the *Message Adapter* of the communication partner to exchange already secured messages. Since the underlying existing systems can only handle messages in an established format, the *Converter* module can convert, i.e., EDIFACT messages to and from messages suitable for the *Creator & Validator.*

When sending an EDIFACT message from A to B, A's message is converted and then the *Creator* module encrypts and signs parts of this message according to a policy. B's message adapter receives the secured message from A's message adapter and applies its *Validator* module. B verifies signatures and decrypts those parts that have been encrypted for B. B also checks if signatures and encryptions have been applied according to the policy. Finally, B converts the validated and partly decrypted message into EDIFACT for B's existing system.

The policy is provided by the *Policy Store.* The complexity and detailed implementation of the *Policy Store* vastly depends on the environment the system is used in, i.e., in an environment with only a few unchanging participants the *Policy Store* might be a static part of the *Message Adapter*. On the other hand, in an environment with many ever-changing participants the *Policy Store* might be its own complex software system which manages the trust relations between the different actors and is adapting the security policy accordingly.

The cryptographic keys, which are necessary to encrypt and sign the various parts of messages, are provided by the *Key Management* module. The implementation of this module also heavily depends on the environment and on the agreements among the participants. Again, in an unchanging environment with a few actors, this module may be static, while in a large system with many different actors the *Key Management* module will be connected to a PKI to provide the public keys of participants. In any case, the *Key Management* module should also provide fresh symmetric keys for the necessary encryptions. The *Creator* module further encrypts symmetric keys with the public key of readers as designated by the policy. These encrypted symmetric keys are attached to the corresponding message.

Apart from encryptions, some parts of a message are signed with the private key of the sender. In addition, some parts of the message have already been signed by previous messages. These signatures are also subject of the policy and are attached to the message and additionally kept in a *Signature Store.* The *Signature Store* is used for forensic purposes and might be accessed in case of policy violations. The implementation of such a module might be either local or system-wide operated by a trusted third party.

### *4.3 Comparing both approaches*

Comparing both approaches, the blockchain solution requires modifications of the current workflow. The blockchain acts as common database storing the unique lifecycle state of a container that determines the further workflow. This approach ensures integrity. Since committed transactions on a blockchain are immutable the whole workflow is also traceable and accountable. A possible weak spot may be confidentiality as all peers have identical copies of the whole blockchain. In fact, confidential data must not be stored in blockchains but separately off-chain. For the given export and import scenario we do not consider container states and attributes to be very sensitive. Furthermore, all peers of the permissioned blockchain are known as part of the consortium, thus misbehaving peers can be sanctioned.

In the peer-to-peer approach it is easier to adhere to confidentiality restrictions as the data is generally sent only to dedicated communication partners. Additionally, there are established approaches to send sensitive data in a secure way via unsecure channels. Concerning integrity, in turn, we require a flexible signature scheme and normalized messages to guarantee (some degree of) accountability. Due to the parallelism in which transactions can take place and due to possible delays in communicating such transactions to others, we can only expect eventual consistency in theory. However, in practice the information transitions in the IT-world are mostly mirrored and synchronized by linearizable activities in the real world.

### 5. Conclusion

In this paper we presented a security architecture developed for an IT-ecosystem for ports. We specified a global access security policy and provided two alternative security mechanisms (and their potential realizations) to enforce this policy in practice. Both proposed solutions rely on a PKI and promise to increase the integrity and confidentiality of the current workflows without compromising availability. In fact, the whole IT-ecosystem is expected to become less vulnerable against cyberattacks.

Our research was based on the investigation of various workflows for exporting and importing container shipments within German ports, whereby the details were provided by various actors involved in these workflows like logistics companies, shipping lines, terminals, port authorities and PCS providers. Currently we extend these approaches in different ways. First, we transfer the presented methods to secure the interaction into a formal specification of a security protocol (and corresponding attacker model) allowing us to use standard verification techniques, like (Thayer et al., 1999), to formally guarantee the required security properties of the considered workflows. Second, we are in discussion with various actors (including some large shipping line worldwide) of how to implement and transfer this approach into practice.